\documentclass[aps,twocolumn,showpacs,prr,amsmath,amssymb,floatfix,superscriptaddress]{revtex4-2}
\usepackage{color}
\usepackage[usenames,dvipsnames]{xcolor}
\usepackage[colorlinks=true,citecolor=Cerulean,linkcolor=RubineRed,urlcolor=Cerulean]{hyperref}
\usepackage{soul,xcolor}
\usepackage{mathptmx} 
\usepackage{dcolumn}
\usepackage{bm}
\usepackage{amsmath}
\usepackage{amssymb}
\usepackage{subfigure}
\usepackage{ulem}
\usepackage{graphicx}
\usepackage{fontenc}
\usepackage{tipa}

\begin{document}

	\title{Correlation-enhanced stability of microscopic cyclic heat engines}
	
	\author{Guo-Hua Xu}
	\email[]{guohuax@zju.edu.cn}
	\affiliation{Department of Physics and Zhejiang Institute of Modern Physics, Zhejiang University, Hangzhou, Zhejiang 310027, China}
	
	\author{Gentaro Watanabe}
	\email[]{gentaro@zju.edu.cn}
	\affiliation{Department of Physics and Zhejiang Institute of Modern Physics, Zhejiang University, Hangzhou, Zhejiang 310027, China}
	\affiliation{Zhejiang Province Key Laboratory of Quantum Technology and Device, Zhejiang University, Hangzhou, Zhejiang 310027, China}
	
	\date{\today}
	
	\begin{abstract}
		For cyclic heat engines operating in a finite cycle period, thermodynamic quantities have intercycle and intracycle correlations. By tuning the driving protocol appropriately, we can get the negative intercycle correlation to reduce the fluctuation of work through multiple cycles, which leads to the enhanced stability compared to the single-cycle operation. Taking the Otto engine with an overdamped Brownian particle as a working substance, we identify a scenario to get such enhanced stability by the intercycle correlation. Furthermore, we demonstrate that the enhancement can be readily realized in the current experiments for a wide range of protocols. By tuning the parameters within the experimentally achievable range, the uncertainty of work can be reduced to below $\sim50 \%$.
	\end{abstract}
	
	\maketitle
	
\textit{Introduction.}
With the advanced technology, various microscopic thermal devices have been fabricated on the submicron scale \cite{Hugel,Steeneken,Shoichi,Blickle_experiment,Brownian_Carnot,Argun,Krishnamurthy,Colloidal_heat_engines,Exp_review,Exp_molecular}. Among them, an important breakthrough for the exploration beyond conventional macroscopic thermodynamics is the experimental realization of the so-called Brownian heat engine \cite{Brownian_Carnot,Blickle_experiment,Krishnamurthy,Argun}, which consists of a Brownian particle subject to a time-dependent optical trap. 
In contrast to conventional macroscopic heat engines, fluctuations of thermodynamic quantities are significant in microscopic heat engines due to the small number of degrees of freedom in their working substance \cite{Noneq_smallsys,Ciliberto_fluct}.
In the past three decades, stochastic thermodynamics has been developed to formulate laws of thermodynamics for fluctuating thermodynamic quantities of small systems, and has had great success in understanding of thermodynamics of small systems \cite{sekimoto_book,Seifert_2012,Seifert_2019,Jarzynski_review}. Motivated by the experimental realization of microscopic heat engines and the theoretical advances in thermodynamics of small systems, there is a surge of activity on the study of microscopic heat engines \cite{SekimotoPRE2000,Schmiedl_2007,HolubecJStatMech2014,Cyclic_Engine,DechantPRL2015,DechantEPL2017,Plata_irr,selfoscillation,FodorEPL2021,BrownianMotor,BrownianGyrator,Trade-OffRelation,Geo_Feature,quan_CarnotatMP,Single-Ion_EMP,activeBrownian}. 
Recently, fluctuations of the performance of microscopic heat engines and characterization of their performance beyond the mean values of thermodynamic quantities have become an active research topic \cite{Sinitsyn_2011,Lahiri_2012,Campisi_2014,Single-particle_HE,Otto_powerlaw,ito2019universal,Saryal_fluct_qOtto,saryal2021universal,Brandner_geo,Miller_geo,watanabe2021finitetime,holubec2021fluctuations,chen2021microscopic,fluct_generalLangevinsys,fluct_periodicallydriven,work_fluct,Work_distribution}.

Nevertheless, many studies of cyclic heat engines so far consider single-cycle operation and focus on the performance within a single cycle. In these studies, fluctuations of the thermodynamic quantities usually include only the intracycle correlation. In the quasistatic limit, since the thermal noise erases the correlation among thermodynamic quantities in different cycles \cite{sekimoto_book,ito2019universal}, it is sufficient to describe the fluctuations focusing on a single cycle. However, to get the nonzero power output, we need to operate engines in a finite cycle period. In this case, the effect of the intercycle correlation becomes non-negligible. Therefore, for engine operations over multiple cycles, assessing the performance within a single cycle is insufficient. Instead, assessments of the engine performance should address the global process over multiple cycles to include intercycle correlations.

Recently, fluctuations including intercycle correlations also started to be discussed. For example, various properties of the stochastic efficiency have been derived \cite{stocastic_efficiency,stocastic_efficiency_2,Stoch_eta_3,SE_4,SE_5,SE_6,SE_7}, and thermodynamic uncertainty relations which give a lower bound of uncertainties of the current \cite{original_TUR,proof_ori_TUR,review_TUR,Pietzonka_Trade-Off,Holubec_CTUR,modified_EP_1,modified_EP_2,Operationally_Accessible} have been generalized for cyclic heat engines in the long-time limit \cite{modified_EP_1,modified_EP_2,Operationally_Accessible}.

However, the role of the time correlation in fluctuations of thermodynamic quantities has not been thoroughly explored. Since engines are supposed to operate over multiple cycles consecutively with a finite cycle period in practical situations, there is a great demand for a scheme to prevent the degradation of performance in multiple cycles by the intercycle correlation effect. In this Letter, by clarifying the effect of time correlation of work in microscopic heat engines with a finite cycle period, we identify such a scheme to reduce the fluctuation of work output. Since the fluctuation of work output is comparable to or even bigger than the average of work output in current experiments of small heat engines \cite{Blickle_experiment,Brownian_Carnot}, reducing the fluctuation of work output is a crucial issue. Taking an example of the Otto engine using a Brownian particle as a working substance, we demonstrate that the reduction of the fluctuation of work output can be realized in a robust manner in the current experiments, and this reduction can be more than $50\%$.

\textit{Setup.}
We study a small cyclic heat engine whose working substance is in contact with a heat bath with the controllable temperature $T(t)$ (we set the Boltzmann constant $k_B = 1$ throughout the Letter). The working substance is described by the Hamiltonian $H(\Gamma,t)$ with an external control parameter $\lambda(t)$, where $\Gamma$ is the microstate of the working substance in the phase space. The engine is driven by time-periodically modulating $T$ and $\lambda$ with period $\tau$, i.e., $T(t)=T(t+\tau)$ and $\lambda(t) = \lambda(t+\tau)$. Under such a protocol, we assume the engine is already driven into a periodic state with the probability distribution function (PDF) satisfying $p(\Gamma,t) = p(\Gamma,t+\tau)$ after running many cycles \cite{Cyclic_Engine}. Therefore, we can represent time $t$ by the phase as $\theta=2\pi t/\tau$, and assign the initial phase $\theta_0$ for the staring point of the cycle. 

The work $W_{\theta_0}^{(n)}$ extracted through $n$ cycles with the initial phase $\theta_0$ is a random variable given by
\begin{equation}
	W_{\theta_0}^{(n)} = - \int_{\theta_0 \tau/2\pi}^{n\tau+\theta_0 \tau/2\pi} \frac{\partial H(\Gamma,t) } {\partial \lambda(t)} \dot{\lambda}(t) \mathrm{d} t,
\end{equation}
where the integral follows the Stratonovich rule \cite{sekimoto_book}. The ensemble average $\langle W^{(n)}_{\theta_0} \rangle$ of work is independent of $\theta_0$, and satisfies $\langle W^{(n)}_{\theta_0} \rangle = n \langle W^{(1)} \rangle$, where $\langle \cdots \rangle = \int \mathcal{D}[\Gamma(t)] p[\Gamma(t)] \cdots$ is the path integral over all the possible trajectories $\Gamma(t)$.

The variance of work with initial time $t_0 = \theta_0 \tau/(2\pi)$ is given by
\begin{equation}
	\text{Var}\big[W_{\theta_0}^{(n)} \big] = \int_{t_0}^{n\tau+t_0} \mathrm{d} t \int_{t_0}^{n\tau+t_0} \mathrm{d} t' ~C(t,t') \label{var},
\end{equation}
where the covariance function of power $\dot{W}\equiv -\partial_{\lambda}H(\Gamma,t)\dot{\lambda}(t)$ is defined as $C(t,t') \equiv \langle\dot{W} (t)\dot{W} (t')\rangle - \langle\dot{W} (t)\rangle\langle\dot{W} (t')\rangle$. The variance $\text{Var}\big[W_{\theta_0}^{(n)} \big]$ of work can be given by the sum of the contribution from each cycle, $n\text{Var}\big[W_{\theta_0}^{(1)} \big]$, and the remaining contribution denoted by $\mathcal{C}_{\theta_0}^{(n)}$:
\begin{align}
	\text{Var}\big[W_{\theta_0}^{(n)} \big] = n\text{Var}\big[W_{\theta_0}^{(1)} \big] + \mathcal{C}_{\theta_0}^{(n)}.
\end{align}
Here, the first term can be identified as the intracycle correlation within each single cycle and the second term $\mathcal{C}_{\theta_0}^{(n)}$ can be regarded as the intercycle correlation between different cycles. Since the system is not in a steady state, $\text{Var}\big[W_{\theta_0}^{(n)} \big]$ changes with $\theta_0$. However, the $\theta_0$-dependence is negligible for $n\to\infty$ because the correlation decays exponentially in time.

In this Letter, we use the single-cycle uncertainty $\Delta^{(1)}_{\theta_0}\equiv{\text{Var} \big[  W_{\theta_0}^{(1)} \big]}/{\langle W_{\theta_0}^{(1)} \rangle^2}$ to describe the fluctuation of work within each single cycle. According to the law of large number, the uncertainty of work extracted through a large numbers $n$ of cycles vanishes as $\sim 1/n$. Therefore, we use the scaled uncertainty $\Delta^{\infty}$ for infinite cycles defined as
\begin{equation}
	\Delta^{\infty} = \lim_{n\to \infty} \Delta^{(n)}_{\theta_0} \equiv 
	\lim_{n\to \infty} n \frac{\text{Var} \big[  W_{\theta_0}^{(n)} \big]}{\langle W_{\theta_0}^{(n)} \rangle^2}. \label{uncertainty_inf}
\end{equation}
Note that the $\theta_0$ dependence of $\Delta^{(n)}_{\theta_0}$ vanishes in the limit of $n\rightarrow\infty$ because $\text{Var} \big[  W_{\theta_0}^{(n)} \big]$ does so and $\langle W_{\theta_0}^{(n)} \rangle$ is independent of $\theta_0$. The multicycle uncertainty $\Delta^{(n)}_{\theta_0} (n\ge2)$ defined in Eq.~\eqref{uncertainty_inf} is the quantity to be compared with $\Delta^{(1)}_{\theta_0}$. For a large cycle period, where the intercycle correlation is negligible, $\mathcal{C}_{\theta_0}^{(n)}\simeq0$, $W_{\theta_0}^{(n)}$ is diffusive with $\text{Var} \big[  W_{\theta_0}^{(n)} \big] = n \, \text{Var} \big[  W_{\theta_0}^{(1)} \big]$, and we get $\Delta^{\infty} =  \Delta^{(1)}_{\theta_0}$. On the other hand, for a small cycle period comparable to the relaxation time of the working substance, the intercycle correlation is significant. Our goal is to find an appropriate protocol which yields $\Delta^{\infty} < \Delta^{(1)}_{\theta_0}$ (i.e., $\mathcal{C}_{\theta_0}^{\infty}<0$) for arbitrary $\theta_0$.

\textit{Relation between the single-cycle and multicycle uncertainties.}
To discuss the relationship between the uncertainties within a single cycle $\Delta^{(1)}_{\theta_0}$ and infinite cycles $\Delta^{\infty}$, we consider an overdamped Brownian particle trapped in a one-dimensional harmonic oscillator potential with the Hamiltonian
\begin{equation}
	H(x,t) = \frac{1}{2} \lambda(t)\, x(t)^2.
\end{equation}
Here, $\lambda(t)$ is the stiffness of the potential which serves as a mechanical control parameter and $x(t)$ is the position of the Brownian particle. This system is described by the Ornstein-Uhlenbeck process \cite{handbook}. The correlation function $\phi(t,t') \equiv \langle x(t)x(t') \rangle$ with $\phi(t,t') = \phi(t',t)$ is derived from  the solution of the It\^o stochastic differential equation for this process \cite{[See Supplemental Material for details]SM}. The resulting correlation function $\phi(t,t')$ for $t<t'$ is given by
\begin{equation}
	\phi(t,t') = \phi(t,t)\exp\bigg[ -\mu\int_{t}^{t'}\mathrm{d}s \lambda(s) \bigg], \label{corr_func} 
\end{equation}
where $\mu$ is  the mobility. 
In addition, since $\phi(t,t)$ is periodic in time, we have 
\begin{equation}
	\phi(t+\tau,t'+\tau)  = \phi(t,t') . \label{phi_tau}
\end{equation}

The covariance function of power becomes
$C(t,t') = \frac{1}{2}\dot{\lambda}(t)\dot{\lambda}(t') \phi(t,t')^2$ \cite{SM}. From Eqs.~\eqref{corr_func} and \eqref{phi_tau}, we get the following properties of the covariance function: $C(t+\tau,t'+\tau)  = C(t,t')$ and $C(t,t'+\tau)  = a C(t,t')$, where $a \equiv \exp[ -2\mu\int_{0}^{\tau}\mathrm{d}t \lambda(t) ]<1$. Therefore, $C(t,t')$ decays exponentially in time when $|t-t'| \gg \tau$, and the correlation time of work is given by $\tau_{\text{corr}} = 2\mu\int_{0}^{\tau}\mathrm{d}t \lambda(t)/\tau$. 

From the above properties of $C(t,t')$, we can write the intercycle correlation $\mathcal{C}_{\theta_0}^{(2)}$ within the two successive cycles as $\mathcal{C}_{\theta_0}^{(2)}=[a+\gamma(\theta_0)] \text{Var}\big[  W_{\theta_0}^{(1)} \big]$, where	
\begin{equation}
	\gamma(\theta_0) \equiv 2 \int_{\tau+t_0}^{2\tau+t_0} \mathrm{d}t' \int_{t'-\tau}^{\tau+t_0} \mathrm{d}t~\frac{ C(t,t') }{\text{Var}\big[  W_{\theta_0}^{(1)} \big]} . \label{gamma}
\end{equation}   
In the same way, we can write $\mathcal{C}_{\theta_0}^{(n)}$ in terms of $a$, $\gamma(\theta_0)$, and  Var$\big[  W_{\theta_0}^{(1)} \big]$ \cite{SM}. Then, the uncertainty for $n$ cycles reads \cite{SM}
\begin{equation}
	n\Delta^{(n)}_{\theta_0} = \bigg[ \left( n - s_n \right) \frac{1 + \gamma(\theta_0)}{1-a} + s_n \bigg]\Delta^{(1)}_{\theta_0} \label{n_Delta_relation} ,
\end{equation}
where $s_n \equiv (1-a^n)/(1-a) \ge 1$.
For infinite cycles, we get
\begin{equation}
	\frac{ \Delta^{(1)}_{\theta_0} }{\Delta^{\infty}} = \frac{1-a}{1+\gamma (\theta_0) } \label{main}.
\end{equation} 
For finite $n$ cycles, the uncertainty derived from Eqs.~\eqref{n_Delta_relation} and \eqref{main} reads
\begin{equation}
	\Delta^{(n)}_{\theta_0} = \bigg( 1 - \frac{s_n}{n} \bigg) \Delta^{\infty} + \frac{s_n}{n} \Delta^{(1)}_{\theta_0}. \label{main2}
\end{equation}

If the intercycle correlation $\mathcal{C}_{\theta_0}^{(2)}$ is negative, i.e., $a+\gamma(\theta_0) < 0$, we get $\Delta^{(1)}_{\theta_0} >\Delta^{(n)}_{\theta_0}> \Delta^{\infty}$ from Eqs.~\eqref{main} and \eqref{main2} \footnote{In this particular model, the sign of $\mathcal{C}_{\theta_0}^{(2)}$ and $\mathcal{C}_{\theta_0}^{\infty}$ are the same, so that the conditions $\mathcal{C}_{\theta_0}^{(2)}<0$ and $\mathcal{C}_{\theta_0}^{\infty}<0$ for $\Delta^{\infty}<\Delta^{(1)}_{\theta_0}$ are equivalent}. This means that the negative covariance of work between two successive cycles, $\mathcal{C}_{\theta_0}^{(2)}<0$, indicates the reduction of uncertainty of work in multiple cycles. It is vice versa for the positive intercycle correlation. It is noted that the essential point of the above discussion is the exponential decay in time of the correlation functions. Even if the effect of inertia is non-negligible beyond the overdamped limit, the correlation functions can still be exponential in time with a smaller correlation time in the overdamped regime \cite{SM}. In addition, in the strongly underdamped regime, the correlation functions can be well approximated by an exponentially decaying function with a large correlation time $\tau_{\text{corr}} \simeq \gamma^{-1}$ obtained by averaging over the rapid oscillation \cite{SM}. Therefore, for the both cases, the above results can still apply, but with a different value of $\tau_{\text{corr}}$.  

It is possible to observe the enhanced stability due to the negative intercycle correlation when $\tau \lesssim \tau_{\text{corr}}$. To show this effect, below we consider a simple Brownian Otto engine, where the analytical result can be obtained. However, a similar result is also obtained for the Carnot cycle \cite{SM}. 

\begin{figure}[t]
	\includegraphics[width=0.6 \columnwidth]{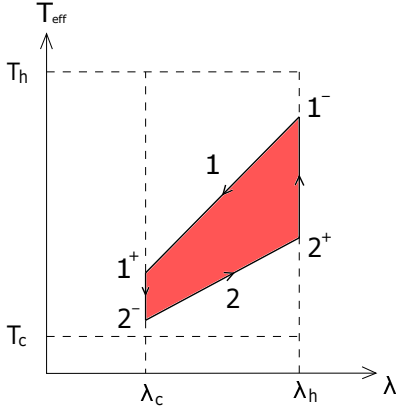}
	\caption{Brownian Otto cycle with a finite cycle period on the $\lambda$-$T_{\text{eff}}$ plane. Stroke $1$ and $2$ are isentropic expansion and compression, respectively. $1^+$ ($1^-$) is the node after (before) the isentropic expansion and $2^+$ ($2^-$) is the node after (before) the isentropic compression. Since the durations of isochoric strokes are finite, the effective temperature at nodes $1^-$ and $2^-$ are different from $T_h$ and $T_c$, respectively.}
	\label{fig:Otto_protocol}
\end{figure}

\textit{Results for the Brownian Otto cycle.}
Next, taking the Brownian Otto engine as an example, we demonstrate that the negative intercycle correlation can be realized in a wide range of parameters in the driving protocol. We still consider an overdamped Brownian particle in a harmonic oscillator potential described by the Ornstein-Uhlenbeck process. In this model, since the PDF $p(x,t)$ of any periodic state is Gaussian, we can define the effective temperature $T_{\text{eff}}$ of the Brownian particle given by $T_{\text{eff}} = \lambda\langle x^2\rangle$ \cite{Schmiedl_2007}. The Brownian Otto engine consists of two isochoric and two isentropic strokes as shown in Fig.~\ref{fig:Otto_protocol} \footnote{In experiments of the Brownian heat engines, adiabatic strokes are often replaced by isentropic strokes \cite{Brownian_Carnot, MartinezPRL2015, Colloidal_heat_engines} since the working substance is always in contact with the environment (water), so that it is impossible to thermally isolate from the environment. In these isentropic strokes, the parameter $\lambda$ and the temperature are controlled to keep the mean value of the entropy of the working substance constant.}. During the hot (cold) isochoric strokes, the temperature $T$ of the bath and the parameter $\lambda$ are fixed at $T_h$ and $\lambda_h$ ($T_c$ and $\lambda_c$), respectively, for the duration $\tau_h$ ($\tau_c$) with $\lambda_c<\lambda_h$. During the isentropic strokes, $T$ and $\lambda$ are quenched simultaneously in a way such that the Shannon entropy $S \equiv-\langle\ln p\rangle$ is unchanged \cite{Schmiedl_2007}. We assume that the isentropic strokes are instantaneous, so that the cycle period is given by $\tau=\tau_h+\tau_c$.

\begin{figure}[t]
	\subfigure{
		\includegraphics[width=0.75 \columnwidth]{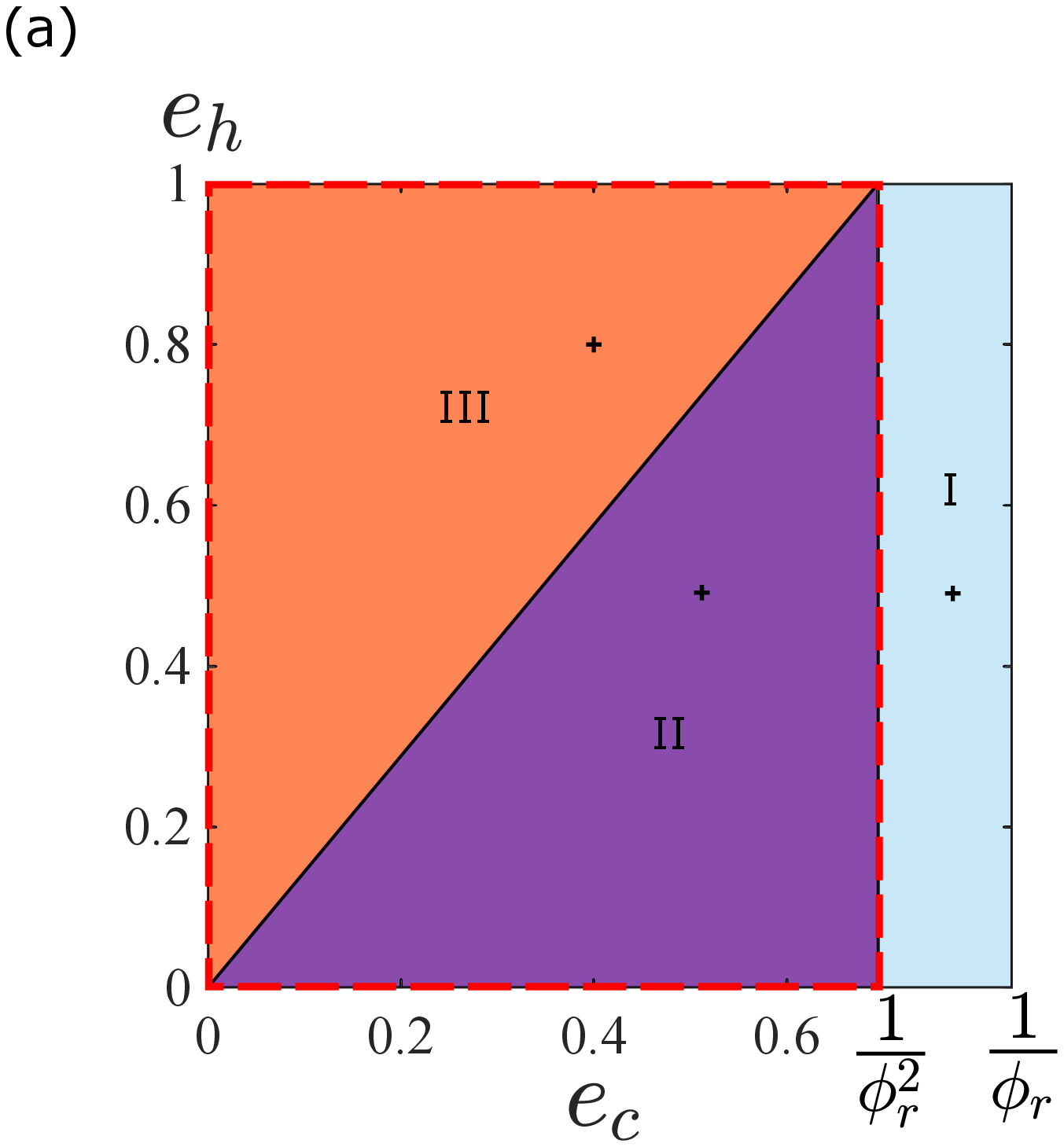} 
		\label{fig:Delta}
	}
	\quad
	\subfigure{
		\includegraphics[width=1 \columnwidth]{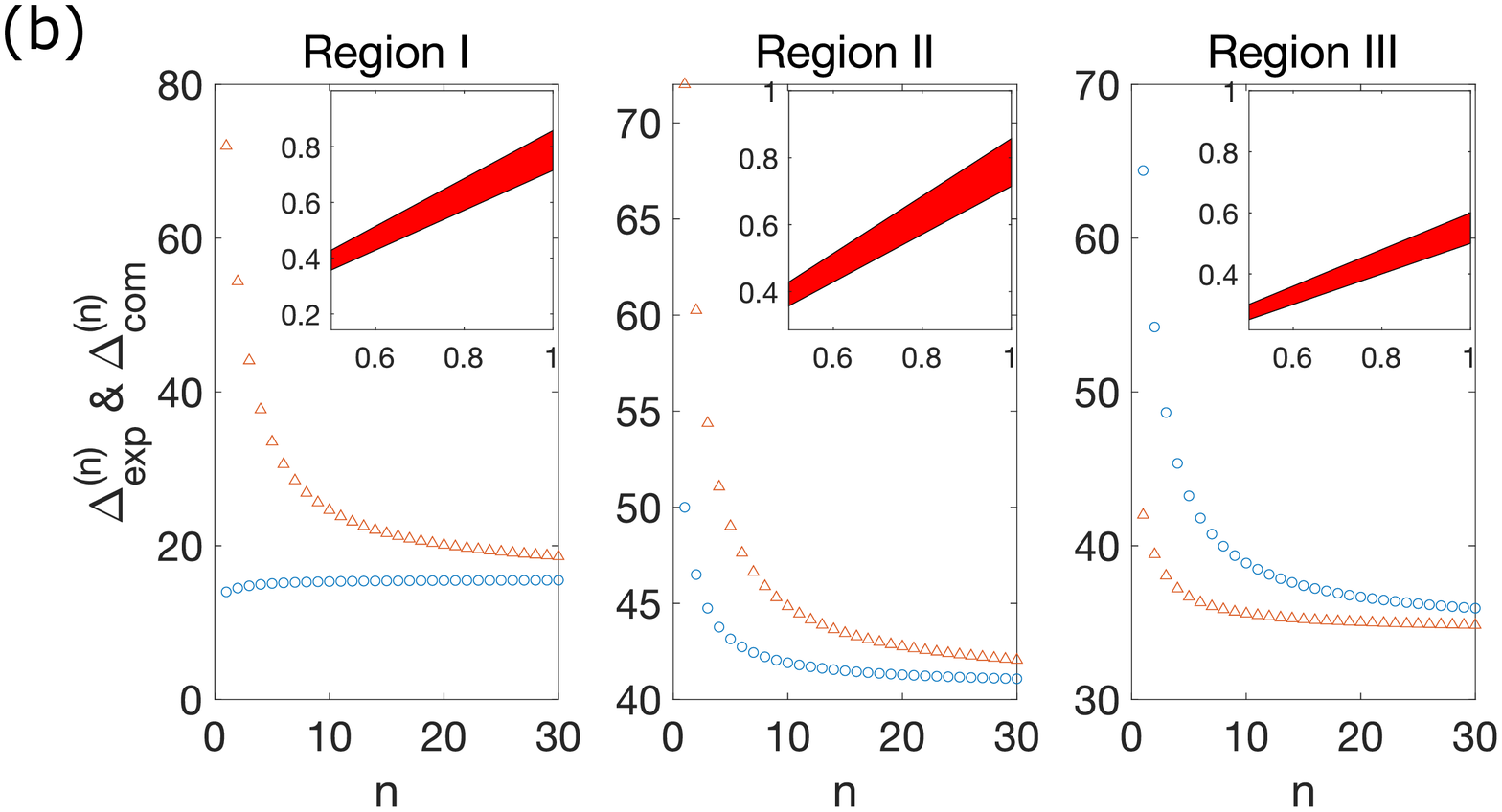}
		\label{fig:Dn}
	}
	\caption{Mapping out the regions of reduced fluctuation by the intercycle correlation. \subref{fig:Delta} The three regions with different order of the uncertainties: $\Delta_{ \text{com} }^{(1)} > \Delta^{\infty} > \Delta_{ \text{exp} }^{(1)}$ in region \uppercase\expandafter{\romannumeral1}, $\Delta_{ \text{com} }^{(1)} > \Delta_{ \text{exp} }^{(1)} > \Delta^{\infty}$ in region \uppercase\expandafter{\romannumeral2}, $\Delta_{ \text{exp} }^{(1)} > \Delta_{ \text{com} }^{(1)} > \Delta^{\infty}$ in region \uppercase\expandafter{\romannumeral3}. The uncertainties are reduced by the intercycle correlation in regions \uppercase\expandafter{\romannumeral2} and \uppercase\expandafter{\romannumeral3}, which are enclosed by the red dashed line. Here we set $\phi_r = 1.2$. \subref{fig:Dn}  $\Delta^{(n)}_{\theta_0}$ as a function of $n$ for a typical point [shown by the cross symbol in \subref{fig:Delta}] of each region. The blue circles show $\Delta^{(n)}_{\text{exp}}$ and the red triangles show $\Delta^{(n)}_{\text{com}}$. Insets of \subref{fig:Dn} show cycle diagrams on the $\lambda$-$T_{\text{eff}}$ plane for each typical point. Here, we set $T_h = 1$, $\lambda_h = 1$, and $\lambda_c = 0.5$ for all the three cycle diagrams.
	}
\end{figure}

For each $m$th cycle, we assign an odd integer $i = 2m-1$ for the isentropic expansion stroke and an even integer $i = 2m$ for the isentropic compression stroke (see the strokes labeled ``1" and ``2" in Fig.~\ref{fig:Otto_protocol} for $m=1$). Since work is done only in the isentropic strokes, the fluctuation of work can take two values depending on whether $\theta_0$ is in the hot or cold isochoric strokes. Therefore, the analysis can be divided into two cases according to the initial phase: the cycle starts before the isentropic expansion or compression. Then, we get the variance of work for the two cases, Var$[W_{\text{exp}}^{(1)}] = \sum_{i,j=1,2}C_{ij}$ and Var$[W_{\text{com}}^{(1)}] = \sum_{i,j=2,3}C_{ij}$, respectively. Here, the subscript $\theta_0$ in $W_{\theta_0}^{(1)}$ is replaced by ``exp" and ``com" for clarity, and $C_{ij}=\frac{1}{2}(\lambda_h-\lambda_c)^2(-1)^{i-j} \phi_{ij}^2$. In this example, the correlation function $\phi_{ij} \equiv \phi(t_i,t_j)$ is analytically solvable \cite{SM}.

From the analytical solution of $\phi_{ij}$, one can find that the uncertainties $\Delta_{ \text{exp} }^{(1)} $, $\Delta_{ \text{com} }^{(1)}$, and $\Delta^{\infty}$ depend on three parameters \cite{SM}: $e_h\equiv\exp(-2\mu \lambda_h \tau_h)$, $e_c\equiv\exp(-2\mu \lambda_c \tau_c)$, and $\phi_r \equiv \phi_{11}/\phi_{22}$. Here, $e_h$ and $e_c$ are measures of the incompleteness of the equilibration in the hot and cold isochoric strokes, respectively, and $\phi_r$ describes the spread of the width of the PDF of the Brownian particle during the hot isochoric strokes. Since we are interested in the heat engine, the mean value of work should be positive, $\langle W^{(1)} \rangle = (\lambda_h-\lambda_c)(\phi_{11}-\phi_{22})/2 > 0$, and thus $\phi_r > 1$. In addition to the condition $\phi_r > 1$, the region of $\phi_r$ is upper bounded as $\phi_r<1/e_c$ because the parameters $e_h$, $e_c$, and $\phi_r$ are constrained by \cite{SM}
\begin{equation}
	\frac{(1-e_h)(1-\phi_r e_c)}{(1-e_c)(\phi_r- e_h)} = R \label{R},
\end{equation}
where $R \equiv T_c\lambda_h/(T_h\lambda_c)$ describes the reversibility with
\begin{equation}
	\eta = 1 - \frac{\lambda_c}{\lambda_h} = 1 - \frac{1}{R} \frac{T_c}{T_h} < \eta_C . \label{eta}
\end{equation}
Since $0<e_c<1$, $0<e_h<1<\phi_r$, and $0<R<1$, we get $\phi_r<1/e_c$ from Eq.~\eqref{R}. 

Figure~\ref{fig:Delta} is a region map showing which of the uncertainties $\Delta_{ \text{exp} }^{(1)} $, $\Delta_{ \text{com} }^{(1)}$, and $\Delta^{\infty}$ is smaller than the others. Regions \uppercase\expandafter{\romannumeral2} and \uppercase\expandafter{\romannumeral3} are of our interest, where the uncertainty $\Delta^{\infty}$ is smaller than those for a single cycle irrespective of the starting point of the cycle. Figure~\ref{fig:Delta} tells that, if the equilibration in the cold isochoric strokes is sufficient with $e_c<1/\phi_r^2$, we can get the reduction of the uncertainty for multiple cycles. It is noted that, to obtain this reduction, only the degree of equilibration in the cold isochoric strokes matters, but not that in the hot isochoric strokes. 

\begin{figure}[t]
	\includegraphics[width=0.66 \columnwidth]{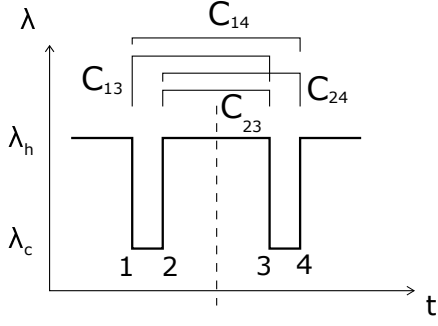}
	\caption{Schematic diagram showing the contributions from the intercycle correlation for the Brownian Otto cycle starting before the isentropic expansion. Strokes 1 and 3 are isentropic expansion and strokes 2 and 4 are isentropic compression. The vertical dashed line represents the boundary between the cycles. 
	}
	\label{fig:correlation}
\end{figure}

\begin{figure}[t]
	\subfigure{
		\includegraphics[width=0.75 \columnwidth]{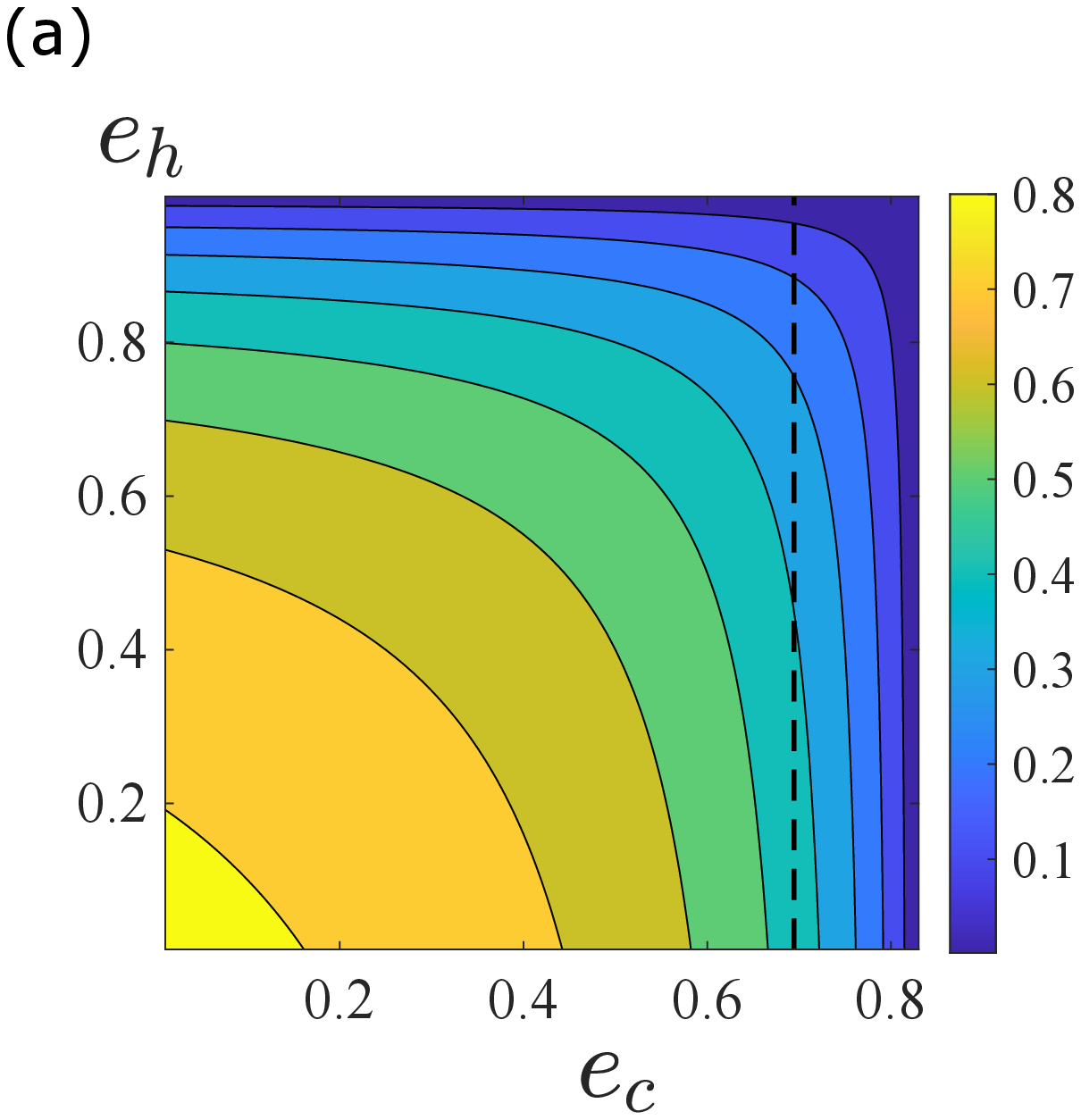} 
		\label{fig:R}
	}
	\quad
	\subfigure{
		\includegraphics[width=1 \columnwidth]{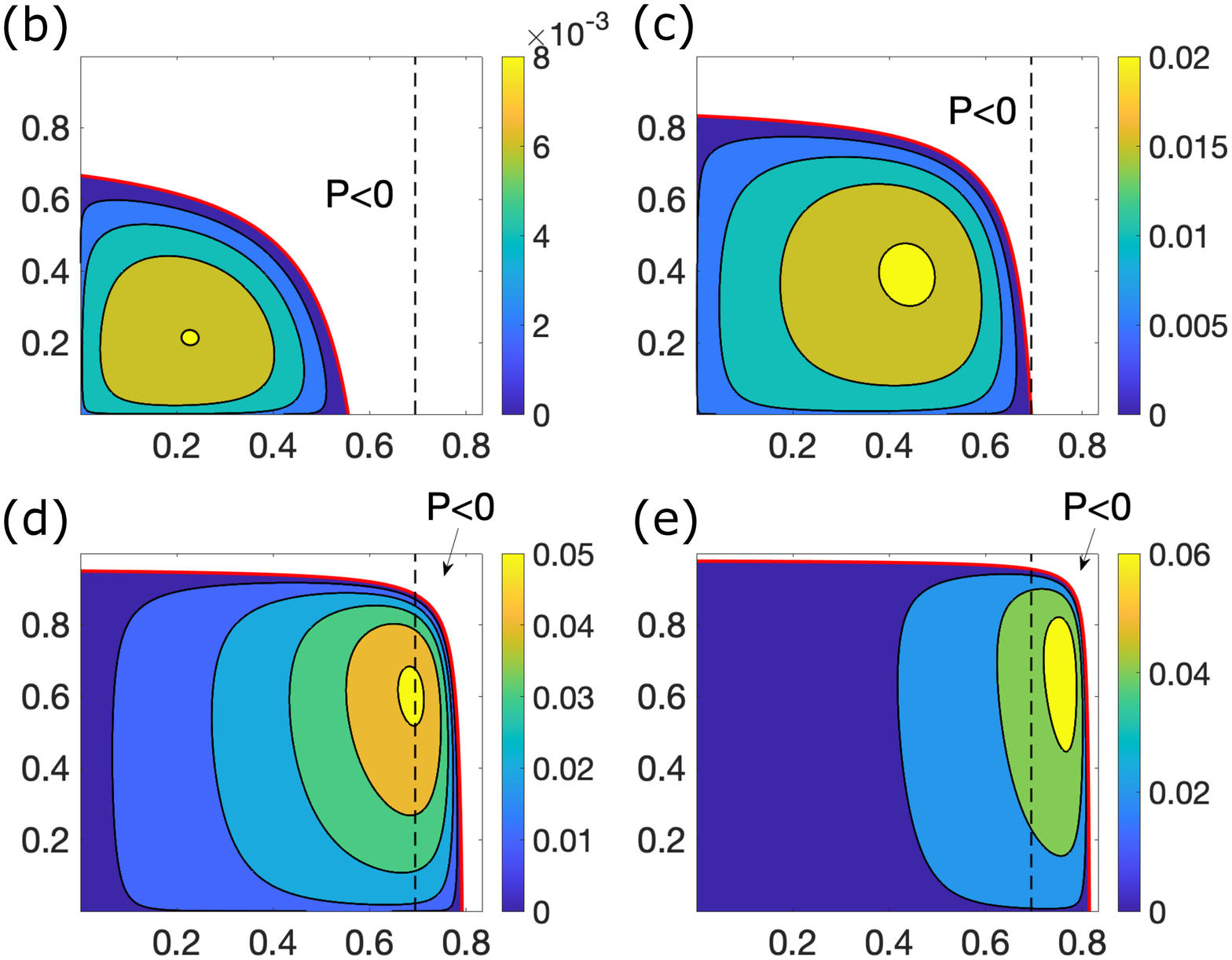}
		\label{fig:Power_1}
	}
	\subfigure{{} \label{fig:Power_2} }
	\subfigure{{} \label{fig:Power_3} }
	\subfigure{{} \label{fig:Power_4} }
	\caption{ \subref{fig:R} Product $R\equiv T_c\lambda_h/(T_h\lambda_c)$ of the compression ratio and the temperature ratio as a function of $e_c$ and $e_h$. \subref{fig:Power_1}-\subref{fig:Power_4} Contour maps of the power as a function of $e_c$ and $e_h$ with given values of $T_h/T_c$, $\lambda_h$, and $\phi_r$. Power is in units of $\mu \lambda_h T_h$. We set $T_h/T_c = 1.6$ [for \subref{fig:Power_1}], $2.2$ [for \subref{fig:Power_2}], $5$ [for \subref{fig:Power_3}], and $10$ [for \subref{fig:Power_4}]. The red solid line shows $P=0$. Since we only focus on the heat engine, values of $P$ for the part with $P<0$ are not shown here. In each figure, the vertical black dashed line shows $e_c=1/\phi_r^2$. Here we set $\phi_r = 1.2$. The contour lines show the values next to the color bar. }
\end{figure}

We can provide a physical understanding of Fig.~\ref{fig:Delta}. An example of the protocol $\lambda(t)$ of the Brownian Otto engine starting before the isentropic expansion (stroke $1$) is shown in Fig.~\ref{fig:correlation}. The intercycle correlation $\mathcal{C}_{\text{exp}}^{(2)}=C_{13}+C_{24}+C_{14}+C_{23}$ is represented by the four lines crossing the boundary between two cycles (vertical dashed line). From Eq.~\eqref{corr_func}, the intercycle correlations in $\mathcal{C}_{\theta_0}^{(2)}$ satisfy $C_{i,i+1} = -e_c C_{ii}$ for odd $i$ and $C_{i,i+1} = -e_h C_{ii}$ for even $i$. The correlation decays with $n$ as $C_{i,j+2n} = a^n C_{ij}$, where $a = e_c e_h < 1$. Therefore, $\mathcal{C}_{\text{exp}}^{(2)}$ is proportional to $C_{11}a - C_{22}e_h \propto \phi_r^2 - 1/e_c$. As we have discussed, the necessary and sufficient condition for the reduction of uncertainty is $\mathcal{C}_{\theta_0}^{(2)}<0$, which gives $e_c<1/\phi_r^2$ corresponding to regions \uppercase\expandafter{\romannumeral2} and \uppercase\expandafter{\romannumeral3}. In the same way, for cycles starting before the isentropic compression, the intercycle correlation $\mathcal{C}_{\text{com}}^{(2)}$ is given by $\mathcal{C}_{\text{com}}^{(2)}\propto 1 - \phi_r^2/e_h$, but it is always smaller than zero. Therefore, we have $\Delta^{\infty}<\Delta_{ \text{com} }^{(1)}$ for arbitrary $e_h$. Summarizing the results for the above two cases, we get $\Delta^{\infty} < \Delta_{ \text{exp} }^{(1)}$ and $\Delta_{ \text{com} }^{(1)}$ provided $e_c < 1/\phi_r^2$. Namely, the fluctuation of work output is reduced in regions \uppercase\expandafter{\romannumeral2} and \uppercase\expandafter{\romannumeral3} for an arbitrary starting point. The difference between regions \uppercase\expandafter{\romannumeral2} and \uppercase\expandafter{\romannumeral3} is in the ordering of $\Delta_{ \text{exp} }^{(1)}$ and $\Delta_{ \text{com} }^{(1)}$ which depends on the intracycle correlation.

\begin{figure}[t]
	\includegraphics[width=1 \columnwidth]{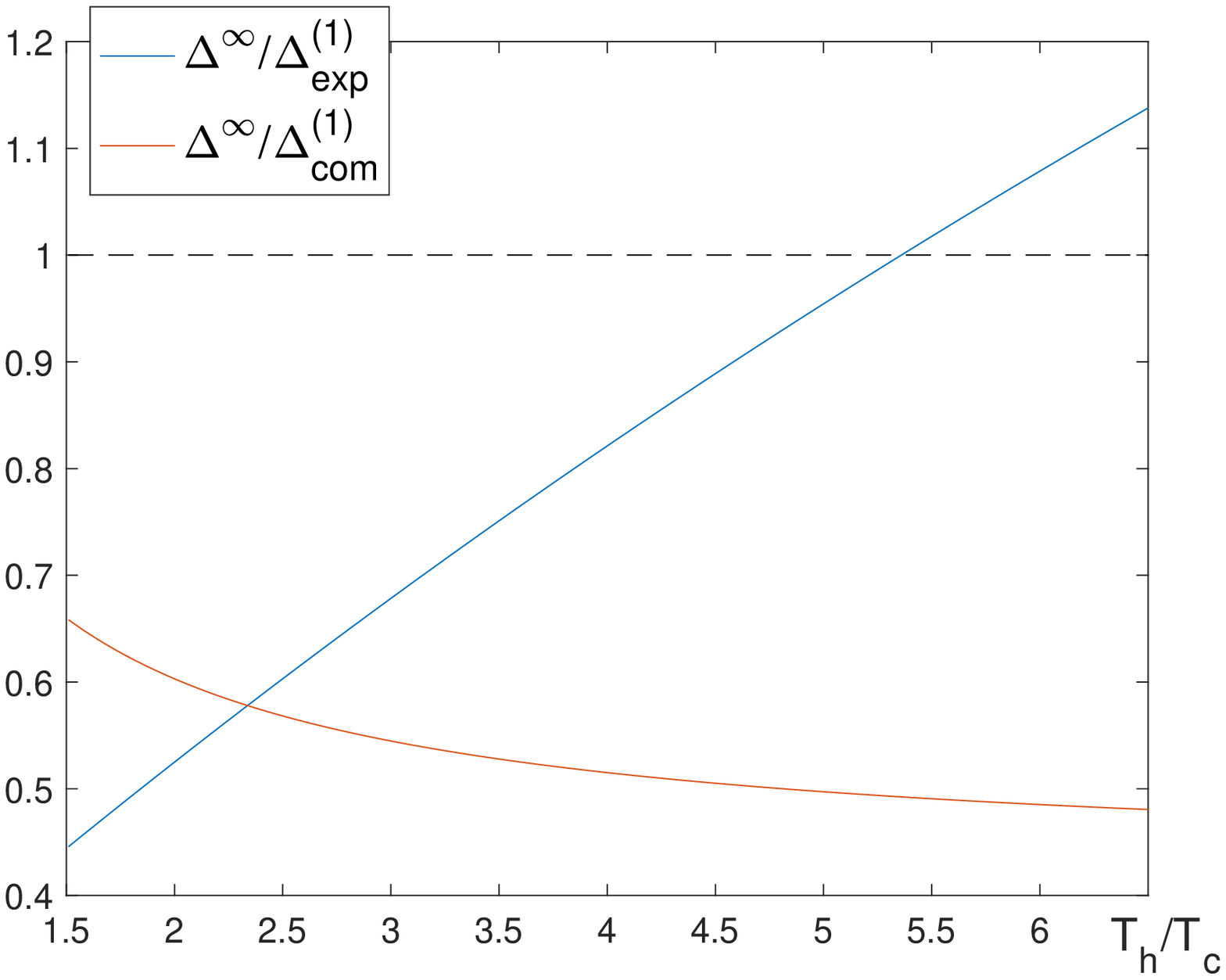}
	\caption{$\Delta^{\infty}/\Delta_{ \text{exp} }^{(1)}$ (blue line) and $\Delta^{\infty}/\Delta_{ \text{com} }^{(1)}$ (red line) as functions of $T_h/T_c$ with $T_c = 300$~K. Here we set $\mu=0.119~\mu$m$\cdot$pN$^{-1}\cdot$ms$^{-1}$ \cite{MestresPRE2014, [{ From the viscosity $\eta$ at room temperature $\eta = 0.89$~pN$\cdot \mu$m$^{-2}$ and the radius of the Brownian particle $r = 0.5$~$\mu$m given by Ref.~\cite{MestresPRE2014}, the mobility $\mu$ is obtained by $\mu = 1/(6\pi\eta r)$ }]mobility}, $\lambda_c=1.6 $~pN$\cdot \mu$m$^{-1}$, $\lambda_h=2.4 $~pN$\cdot \mu$m$^{-1}$, $\tau_c=0.7$~ms, and $\tau_h=0.3$~ms. These parameters are achievable in the current experiment of Ref.~\cite{Brownian_Carnot}. The ratio $\Delta^{\infty}/\Delta_{ \text{exp} }^{(1)}$ is still less than unity even at higher $T_h$ beyond $T_h/T_c=2$. In addition, the ratios $\Delta^{\infty}/\Delta_{ \text{exp} }^{(1)}$ and $\Delta^{\infty}/\Delta_{ \text{com} }^{(1)}$ can reach $\alt 50\%$. }
	\label{fig:Th}
\end{figure}

Finally, we discuss the role of the temperature of the bath and the experimental feasibility to get the reduction of the fluctuation by the intercycle correlation. First, we consider the mean value of the power $P$. As obtained in Ref.~\cite{xu_eta2}, $P$ depends on six parameters: $T_h$, $T_c$, $\lambda_h$, $\lambda_c$, $\tau_h$, and $\tau_c$ \cite{SM}. At any point in the region of $0<e_c<1$ and $0<e_h<1$, the power can be set to any positive value for a given $\phi_r$ by tuning the remaining free parameters, such as $T_h$, $T_c$, and $\lambda_h$. Figures~\ref{fig:Power_1}--\ref{fig:Power_4} show the power for different values of $T_h/T_c$. It can be seen that the point in the $e_c$-$e_h$ plane giving the maximum power can be located in region \uppercase\expandafter{\romannumeral1} or \uppercase\expandafter{\romannumeral2} by tuning $T_h/T_c$. It is noted that we have $R > T_c/T_h$ for the Otto engine with $P>0$ ($\eta>0$) from Eq.~\eqref{eta}. Second, we discuss the role of the temperature ratio $T_h/T_c$ in the correlation-enhanced stability. Figure~\ref{fig:R} shows a contour plot of $R$ as a function of $e_c$ and $e_h$ for a fixed value of $\phi_r$.  As can be seen from Fig.~\ref{fig:R}, if $R$ is larger than that at $e_c=1/\phi_r^2$ and $e_h=0$, it is guaranteed that we are in either region \uppercase\expandafter{\romannumeral2} or \uppercase\expandafter{\romannumeral3}. From Eq.~\eqref{R}, we find that this condition is $R > 1/(\phi_r + 1)$, or
\begin{equation}
	\frac{T_c}{T_h} > \frac{\lambda_c / \lambda_h}{\phi_r + 1}.
\end{equation}
A sufficient condition to satisfy this inequality is $T_h/T_c < 2$, which is easy to realize in experiments. In experiments of microscopic heat engines with Brownian particles \cite{Blickle_experiment,Brownian_Carnot,Argun,Colloidal_heat_engines,Exp_review}, one of the heat bath temperatures (commonly $T_c$) is usually set to be the room temperature: $T_c \sim 300~$K. In such a case, if $T_h$ is $300~\text{K}<T_h < 600~\text{K}$ which is indeed the case in typical experiments \cite{Blickle_experiment,Brownian_Carnot}, it is guaranteed that the fluctuation of work in the Brownian Otto cycle is always reduced for multiple cycles irrespective of the other parameters. To demonstrate the large reduction of $\Delta^{\infty}$ by the intercycle correlation, we plot $\Delta^{\infty}/\Delta_{ \text{exp} }^{(1)}$ and $\Delta^{\infty}/\Delta_{ \text{com} }^{(1)}$ as functions of $T_h/T_c$ in Fig.~\ref{fig:Th} for parameter values accessible in current experiments. Since the work output is zero at $T_h/T_c=\lambda_h/\lambda_c$ and increases with $T_h/T_c$, the region of $T_h/T_c$ shown in Fig.~\ref{fig:Th} gives positive work output. It is noted that, compared to the above-mentioned sufficient condition, $T_h/T_c < 2$, for $\Delta^{\infty} < \Delta_{ \text{exp} }^{(1)}$ and $\Delta_{ \text{com} }^{(1)}$, we can obtain this reduction of $\Delta^{\infty}$ in a much wider temperature region of $T_h/T_c \alt 5.4$. Furthermore, the reduction of $\Delta^{\infty}$ over the single-cycle uncertainties can be very large by appropriately tuning the parameters and protocol. At $T_h/T_c \simeq 2.3$ where the red and blue lines cross, we have the same reduction rate for an arbitrary starting point. In this case, the uncertainty $\Delta^{\infty}$ can be reduced to less than $60\%$ of the single-cycle uncertainties. If we set the starting point of the cycle before the isentropic compression stroke (i.e., the case of the red line), $\Delta^{\infty}$ can be reduced to even below $50\%$ of the single-cycle uncertainty.

\textit{Conclusion.}
Our work has clarified the consequences of time correlation of work over different cycles in cyclic heat engines. If the cycle period is finite, focusing on one cycle is insufficient to discuss fluctuations of the performance of the microscopic heat engines. In particular, taking advantage of the intercycle correlation, the stability of the work output for the multicycle operation can be improved over the single-cycle one. Since such an improvement can be realized in a wide range of protocols, one can further optimize the other performance of the engine (such as efficiency, power, and uncertainty within each cycle). Furthermore, we have demonstrated that our findings can be readily realized in the current experiments. By tuning the parameters within the experimentally achievable range, the uncertainty of work output for infinite cycles can be reduced to less than $50\%$ of the uncertainty for each single cycle. Since the fluctuation of work output can be even larger than the average of the work output in the current experiments \cite{Blickle_experiment,Brownian_Carnot}, our result should provide an important step toward the realization of microscopic heat engines for practical use. The effect of time correlation in other kinds of heat engines, such as autonomous heat engines and self-oscillating heat engines \cite{selfoscillation}, is an interesting future problem. 

\begin{acknowledgements}
	G.~W. is supported by NSF of China (Grant No. 11975199), by the Zhejiang Provincial Natural Science Foundation Key Project (Grant No. LZ19A050001), and by the Zhejiang University 100 Plan. 
\end{acknowledgements}

\bibliography{Corr_effect_ref}

\clearpage
\onecolumngrid

\begin{center}
	
	\newcommand{\beginsupplement}{%
		\setcounter{table}{0}
		\renewcommand{\thetable}{S\arabic{table}}%
		\setcounter{figure}{0}
		\renewcommand{\thefigure}{S\arabic{figure}}%
	}
	
	\textbf{\large --- Supplemental Material --- \\Correlation-enhanced Stability of Microscopic Cyclic Heat Engines}
\end{center}
\newcommand{\beginsupplement}{%
	\setcounter{table}{0}
	\renewcommand{\thetable}{S\arabic{table}}%
	\setcounter{figure}{0}
	\renewcommand{\thefigure}{S\arabic{figure}}%
}

\setcounter{equation}{0}
\setcounter{figure}{0}
\setcounter{table}{0}
\setcounter{page}{1}
\makeatletter
\renewcommand{\theequation}{S\arabic{equation}}
\renewcommand{\thefigure}{S\arabic{figure}}
\renewcommand{\bibnumfmt}[1]{[S#1]}
\renewcommand{\citenumfont}[1]{S#1}
\vspace{0.8 in}

\renewcommand{\bibliography}

\newcommand{\avg}[1]{\ensuremath{\langle #1 \rangle}}
\newcommand{\ve}[1]{\bm{#1}}


\vspace{-2 cm}

\subsection{A. Derivation of the correlation function, Eq.~(6)}

Motion of a Brownian particle in a one-dimensional (1D) harmonic oscillator potential is a Gaussian process $x(t)$ described by the following It\^o stochastic differential equation \cite{Gardiner_suppl}:
\begin{equation}
	\mathrm{d}x(t) = -\mu\lambda(t)\, x(t)\, \mathrm{d}t + \sqrt{2\mu T(t)}\, \mathrm{d}W(t), \label{OU-SDE}
\end{equation}
where $\mathrm{d}W$ is the Wiener noise. The solution of Eq.~\eqref{OU-SDE} is
\begin{equation}
	x(t) = e^{-f(t)} x(0) + \int_0^{t} e^{-f(t)+f(s)} \sqrt{2\mu T(s)}\, \mathrm{d}W(s) \label{eq:x_suppl}
\end{equation}
with $f(t) \equiv \int_0^{t}\mu \lambda(s)\, \mathrm{d}s$. Thus the mean value $\langle x(t) \rangle$ is given by
\begin{equation}
	\langle x(t) \rangle = e^{-f(t)} \langle x(0) \rangle.\label{eq:xavg_suppl}
\end{equation}
Because of the periodicity of the phase-space distribution function of the Brownian particle with the cycle period $\tau$, we have $\avg{x(\tau)} = \avg{x(0)}$. Thus, together with Eq.~(\ref{eq:xavg_suppl}), we get $\avg{x(\tau)} = e^{-f(\tau)} \avg{x(0)} = \avg{x(0)}$, which leads to $\avg{x(0)} = 0$ since $f(\tau) \neq 0$. Therefore, the variance of $x(t)$ becomes $\text{Var}[x(t)] \equiv \langle( x(t) - \langle x(t) \rangle)^2 \rangle = \langle x^2(t) \rangle $. 

The correlation function $\phi(t,t')$ is given by
\begin{equation}
	\phi(t,t') \equiv \langle x(t)\, x(t') \rangle = e^{-[f(t)+f(t')]}  \langle x^2(0) \rangle  + \int_0^{\min(t,\, t')}e^{-[f(t)+f(t')-2f(s)]}\, 2\mu T(s)\, \mathrm{d}s.
\end{equation}
From the periodicity, $\langle x^2(\tau) \rangle = \langle x^2(0) \rangle$, and Eq.~(\ref{eq:x_suppl}), we have
\begin{equation}
	\langle x^2(0) \rangle  = \dfrac{\displaystyle 2\mu \int_0^{\tau}e^{2f(t)}T(t)\, \mathrm{d}t }{e^{2f(\tau)}-1}.
\end{equation}
Therefore, the correlation function $\phi(t,t')$ reads
\begin{equation}
	\phi(t,t') = e^{-[f(t)+f(t')]}\, 2\mu \left[ \int_0^{\min(t,\, t')} e^{2f(s)}T(s)\, \mathrm{d}s + \frac{\displaystyle \int_0^{\tau}e^{2f(s)}T(s)\, \mathrm{d}s}{e^{2f(\tau)}-1} \right] . \label{phi_suppl}
\end{equation}
For $t<t'$, $\phi(t,t')$ satisfies
\begin{equation}
	\phi(t,t') = \phi(t,t)\, \exp\bigg[ -\mu\int_{t}^{t'}\mathrm{d}s\, \lambda(s) \bigg]. \label{phi_simp}
\end{equation}

\subsubsection{ The effect of inertia}

As can be seen from Eq.~\eqref{phi_simp}, the correlation function decays exponentially in the overdamped regime. Our results, Eqs.~$(10)$ and $(11)$ of the main paper, are obtained from such exponentially decaying correlation function. Here, we show that, even if the inertia is non-negligible, the correlation functions can still be exponential in time in the strongly overdamped regime with $\gamma\gg\omega$, where $\omega=\sqrt{\lambda/m}$ is the frequency of the harmonic oscillator trapping potential and $m\gamma=\mu^{-1}$ is the frictional coefficient. In addition, in the strongly underdamped regime with $\gamma\ll \omega$, the correlation function can be well approximated by an exponentially decaying function in time obtained by averaging over the rapid oscillation. Therefore, for the both cases, the similar argument in the main paper can still apply with the effect of inertia, but $\tau_{\text{corr}}$ becomes smaller (larger) in the strongly overdamped (underdamped) regime compared to that in the overdamped limit. As a result, it is harder (easier) to observe the correlation-enhanced stability in the strongly overdamped (underdamped) regime because of the necessary condition: $\tau \lesssim \tau_{\text{corr}}$. 

From the underdamped Langevin equation:
\begin{align}
	m\dot{v} &= -m\gamma v - m \omega^2 x + \sqrt{2m\gamma T} \xi , \\
	\dot{x} &= v ,
\end{align}
with $\xi=\mathrm{d} W(t)/\mathrm{d} t$, we can obtain the two-point correlation function of the positions of the Brownian particle \cite{Frim}:
\begin{equation}
	\phi(0,t)=\langle x(t) x(0) \rangle = \frac{ T}{m\omega^2} \frac{\Lambda_{+}\exp(-\Lambda_{-}t) - \Lambda_{-}\exp(-\Lambda_{+}t)}{\Lambda_{+} - \Lambda_{-}}  \label{phi_over}
\end{equation}
with 
\begin{equation}
	\Lambda_{\pm} = \frac{\gamma}{2} \pm \sqrt{\left(\frac{\gamma}{2}\right)^2 - \omega^2} .
\end{equation}
In the overdamped case with $\gamma \gg \omega$, we have $\Lambda_{+} \gg \Lambda_{-}$ leading to 
\begin{equation}
	\phi(0,t) =  \frac{ T}{m\omega^2} \frac{\exp(-\Lambda_{-}t)}{1-\frac{\Lambda_{-}}{\Lambda_{+}}}. \label{gen}
\end{equation}
Therefore, the correlation function still shows exponential decay. Expand $\Lambda_{-}$ to the second order of $\omega/\gamma$, we have $\Lambda_{-} \simeq \mu\lambda [1+(\omega/\gamma)^2] > \mu\lambda$, where $\lambda=m\omega^2$ and $\mu=(m\gamma)^{-1}$.
It is noted that the correlation time, $\Lambda_{-}^{-1}$, is reduced from that in the overdamped limit, $(\mu \lambda)^{-1}$, due to the effect of inertia. 

We assume that $\lambda$ changes in a timescale $\lambda/\dot{\lambda}$ much larger than $2\pi/\omega$, and the correlation function oscillates with period $2\pi/\omega$ much smaller than $\tau$ in the strongly underdamped regime with $\gamma \ll \omega$ \cite{underdamp}. Averaging over the coarse-grained timescale, which is sufficiently larger than $2\pi/\omega$ but sufficiently smaller than $\tau$, the correlation function of work is given by 
\begin{equation}
	\overline{C(t,t')} = \frac{1}{2}\dot{\lambda}(t)\dot{\lambda}(t') \overline{\phi(t,t')^2},
\end{equation}
where ``-----'' means the average over the coarse-grained timescale.
In the timescale much smaller than $\lambda/\dot{\lambda}$, where the change of $\lambda\equiv m\omega^2$ is negligible, we have the following expression from Eq.~\eqref{phi_over}:
\begin{align}
	\phi(0,t)^2 & = \left( \frac{T}{m\omega^2} \frac{\Lambda_{+}\exp(-\Lambda_{-}t) \Lambda_{-}\exp(-\Lambda_{+}t)}{\Lambda_{+} - \Lambda_{-}} \right)^2 \notag \\
	& = \left( \frac{T}{m\omega^2} \right)^2 \left( \frac{\gamma}{2\kappa}\sin\kappa t + \cos\kappa t \right)^2 \exp(-\gamma t) 
\end{align}
with
\begin{equation}
	\kappa = \omega \sqrt{1-\frac{1}{4}\left(\frac{\gamma}{\omega}\right)^2} = \omega + {O}[(\gamma/ \omega)^2].
\end{equation}
Averaging over each oscillation period, we obtain
\begin{equation}
	\overline{\phi(0,t)^2} \propto \exp(-\gamma t).
\end{equation}
Therefore, also in the strongly underdamped regime of $\gamma\ll\omega$, the correlation function of work is well approximated by an exponentially decaying function in time with the large correlation time $\tau_{\text{corr}}=\gamma^{-1}$.

\subsection{B. Derivation of Eq.~(9)}

\begin{figure}[t]
	\includegraphics[width=0.5 \columnwidth]{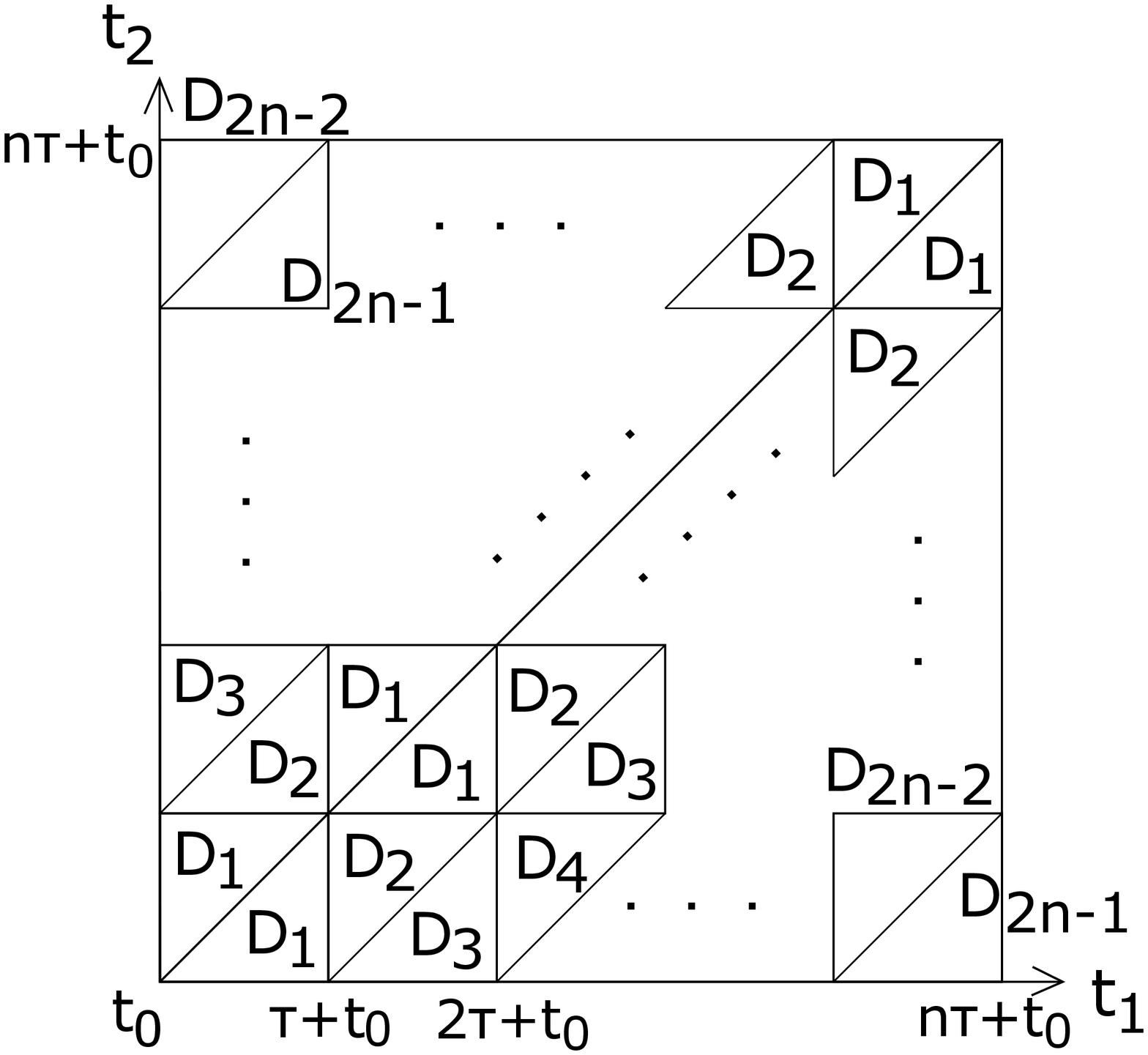}
	\caption{Domain of the time-integration in the variance $\text{Var}\big[W_{\theta_0}^{(n)} \big]$ of work given by Eq.~(\ref{eq:var_suppl}). The horizontal (vertical) axis shows the integration variable $t_1$ ($t_2$) running from $t_0$ to $n\tau+t_0$. The whole domain of the integration is divided by $2n^2$ subdomains $D_i$ with $i=1$, $2$, $\cdots$, $2n-1$. Among these subdomains, equivalent ones are denoted by $D_i$ with the same integer $i$. 
	}
	\label{fig:domain}
\end{figure}

Here, we give $\mathcal{C}_{\theta_0}^{(n)}$ in terms of $a$, $\gamma(\theta_0)$, and  Var$\big[  W_{\theta_0}^{(1)} \big]$. Recall that the variance $\text{Var}\big[W_{\theta_0}^{(n)} \big]$ of work is given by
\begin{align}
	\text{Var}\big[W_{\theta_0}^{(n)} \big] = \int_{t_0}^{n\tau+t_0} \mathrm{d} t_1 \int_{t_0}^{n\tau+t_0} \mathrm{d} t_2\, ~C(t_1,t_2) \label{eq:var_suppl},
\end{align}
where the covariance function $C(t_1,\, t_2) \equiv 2^{-1} \dot{\lambda}(t_1)\, \dot{\lambda}(t_2)\, \phi(t_1,\, t_2)^2$ satisfies the following properties:
$C(t_1+\tau,\, t_2+\tau)  = C(t_1,\, t_2)$ and $C(t_1,\, t_2+\tau)  = a\, C(t_1,\, t_2)$ with $a \equiv \exp[ -2\mu\int_{0}^{\tau}\mathrm{d}t\, \lambda(t) ]$. The domain of the time-integration in the rhs of Eq.~(\ref{eq:var_suppl}) is shown in Fig.~\ref{fig:domain}. In accordance with the periodicity of $C(t_1,\, t_2)$, $C(t_1+\tau,t_2+\tau)  = C(t_1,t_2)$, equivalent subdomains are denoted by $D_i$ with the same integer $i$. Now we introduce 
\begin{equation}
	d_i \equiv \int_{D_i} \mathrm{d} t_1 \mathrm{d} t_2\, C(t_1,t_2),
\end{equation}
and we have
\begin{equation}
	d_1=  \text{Var}\big[W_{\theta_0}^{(1)} \big] /2,
\end{equation}
\begin{equation}
	d_2 = \text{Var}\big[W_{\theta_0}^{(1)} \big]\, \gamma(\theta_0)/2,
\end{equation}
where
\begin{equation}
	\gamma(\theta_0) \equiv 2\int_{\tau+t_0}^{2\tau+t_0} \mathrm{d}t_2 \int_{t_2-\tau}^{\tau+t_0} \mathrm{d}t_1~\frac{ C(t_1,t_2) }{\text{Var}\big[  W_{\theta_0}^{(1)} \big]} , \label{gamma_suppl}
\end{equation}
and 
\begin{equation}
	d_{i+2} = a\cdot d_i
\end{equation}
for $i\ge 2$.

For $n$ cycles, the variance of work is given by
\begin{align}
	\text{Var}\big[W_{\theta_0}^{(n)} \big] & = 2\sum_{k=0}^{n-1} (n-k)\, d_{2k+1} +  2\sum_{k=1}^{n-1} ( n-k )\, d_{2k} \nonumber\\
	& = \bigg[ \bigg( n - s_n \bigg) \frac{1 + \gamma(\theta_0)}{1-a} + s_n \bigg] \text{Var}\big[W_{\theta_0}^{(1)} \big] . \label{varn_suppl}
\end{align}
Here, we have used the following summation formula of the series:
\begin{equation}
	\sum_{k=0}^{n-1} (n-k)\, a^k = \frac{n-a \cdot s_n}{1-a}
\end{equation}
with $s_n \equiv (1-a^n)/(1-a)$. 
From Eq.~(\ref{varn_suppl}), the uncertainty for $n$ cycles reads
\begin{equation}
	n\Delta^{(n)}_{\theta_0} = \bigg[ \bigg( n - s_n \bigg) \frac{1 + \gamma(\theta_0)}{1-a} + s_n \bigg]\Delta^{(1)}_{\theta_0} .
\end{equation}
From Eq.~(\ref{varn_suppl}), we can readily identify $\mathcal{C}_{\theta_0}^{(n)}$ as
\begin{equation}
	\mathcal{C}_{\theta_0}^{(n)} = \frac{n-s_n}{1-a} \, [a+\gamma(\theta_0)]\, \text{Var}\big[  W_{\theta_0}^{(1)} \big].
\end{equation}

\subsection{C: Covariance function of power for an overdamped Brownian particle trapped in a one-dimensional harmonic oscillator potential}

For a Gaussian process, $x(t)$, all the higher-order correlation functions can be decomposed into products of two-point correlation functions using Wick's theorem \cite{Kubo}. For example, we have
\begin{equation}
	\langle x(t)^2 x(t')^2 \rangle = \langle x(t)^2 \rangle \langle x(t')^2 \rangle + 2\langle x(t) x(t') \rangle^2.
\end{equation}
Therefore, for an overdamped Brownian particle trapped in a one-dimensional harmonic oscillator potential with the Hamiltonian, $H(x,t) = \frac{1}{2} \lambda(t)\, x(t)^2$, the covariance function of power is given by
\begin{align}
	C(t,t') &= \langle\dot{W} (t)\dot{W} (t')\rangle - \langle\dot{W} (t)\rangle\langle\dot{W} (t')\rangle \notag \\
	& = \frac{1}{4} \dot{\lambda}(t) \dot{\lambda}(t') ( \langle x(t)^2 x(t')^2 \rangle - \langle x(t)^2 \rangle \langle x(t')^2 \rangle ) \notag \\
	& = \frac{1}{2}\dot{\lambda}(t)\dot{\lambda}(t') \phi(t,t')^2.
\end{align}

\subsection{D: Proof that $\Delta_{ \text{exp} }^{(1)} $, $\Delta_{ \text{com} }^{(1)}$, and $\Delta^{\infty}$ depend on three parameters: $e_h$, $e_c$, and $\phi_r$}

Here, we consider the Brownian Otto engine with the protocol defined in the main paper, and show that the uncertainties $\Delta_{ \text{exp} }^{(1)} $, $\Delta_{ \text{com} }^{(1)}$, and $\Delta^{\infty}$ depend on $e_h$, $e_c$, and $\phi_r$. First, let us consider the case starting before an isentropic expansion stroke. The mean value $\langle W^{(1)} \rangle$ of work is given by 
\begin{equation}
	\langle W^{(1)} \rangle = \frac{1}{2} (\lambda_h-\lambda_c) (\phi_{11} - \phi_{22})
\end{equation}
with $\phi_{ij} \equiv \phi(t_i,\, t_j)$, where $\lambda_h$ and $\lambda_c$ are the stiffness of the potential during the hot and cold isochoric stroke, respectively.
The variance $\text{Var}\big[W_{\text{exp}}^{(1)} \big]$ of work is given by
\begin{equation}
	\text{Var}\big[W_{\text{exp}}^{(1)} \big] = \sum_{i,j = 1,2} C_{ij}
\end{equation}
with $C_{ij} \equiv C(t_i,\, t_j)$. Since $C_{ij} = 2^{-1} (\lambda_h-\lambda_c)^2(-1)^{i-j} \phi_{ij}^2$ and $\phi_{12}^2 = e_c \phi_{11}^2$ with $e_c\equiv\exp(-2\mu \lambda_c \tau_c)$ and $\tau_c$ being the duration of the cold isochoric stroke, the uncertainty $\Delta_{ \text{exp} }^{(1)} \equiv \text{Var}\big[W_{\text{exp}}^{(1)} \big] / \langle W^{(1)} \rangle^2$ is given by
\begin{equation}
	\Delta_{ \text{exp} }^{(1)} = 2\frac{(1-2e_c)\, \phi_r^2 + 1}{(\phi_r-1)^2} \label{D1_exp_suppl}
\end{equation}
with $\phi_r \equiv \phi_{11}/\phi_{22}>1$. Similarly, in the case of starting before an isentropic compression stroke, the variance $\text{Var}\big[W_{\text{com}}^{(1)} \big]$ of work is given by
\begin{equation}
	\text{Var}\big[W_{\text{com}}^{(1)} \big] = \sum_{i,j = 2,3} C_{ij}.
\end{equation}
Since we have $\phi_{23}^2 = e_h \phi_{22}^2$ with $e_h\equiv\exp(-2\mu \lambda_h \tau_h)$ and $\tau_h$ being the duration of the hot isochoric stroke, the uncertainty $\Delta_{ \text{com} }^{(1)}$ is given by
\begin{equation}
	\Delta_{ \text{com} }^{(1)} = 2\frac{\phi_r^2 + (1-2e_h)}{(\phi_r-1)^2}. \label{D1_com_suppl}
\end{equation}

\begin{figure}[t!]
	\includegraphics[width=0.5 \columnwidth]{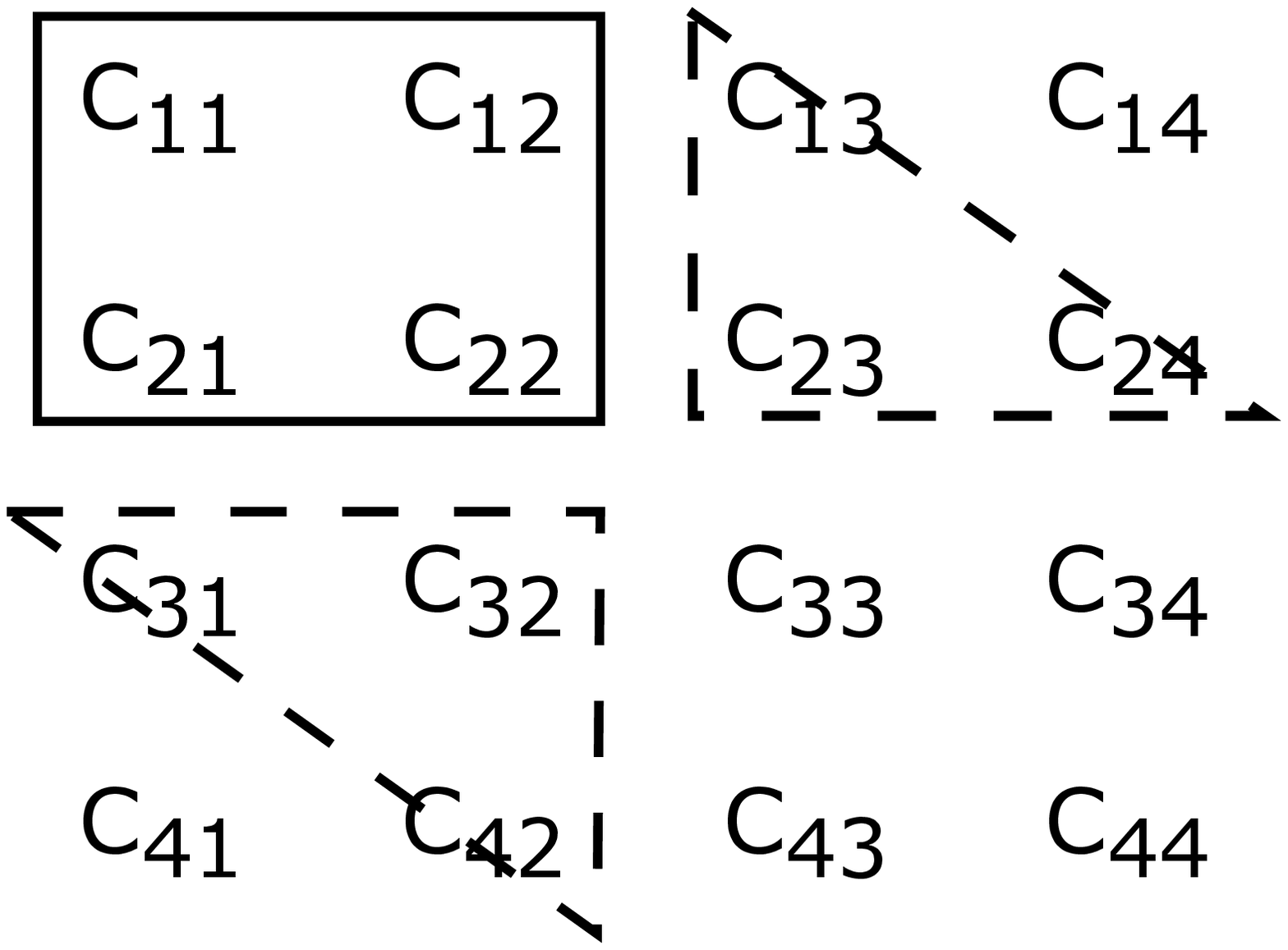}
	\caption{Covariance matrix $C_{ij}$ for two successive cycles starting before an isentropic expansion stroke. The sum of elements $C_{ij}$ enclosed by the solid line equals $\text{Var}\big[W_{\text{exp}}^{(1)} \big]$, and the elements $C_{ij}$ enclosed by the dashed lines contribute to $\gamma_{\text{exp}}$. }
	\label{fig:C}
\end{figure}

Finally, we discuss $\Delta^\infty$. Since $\Delta^\infty$ is given by $\Delta^\infty = [1+\gamma(\theta_0)] (1-a)^{-1} \Delta_{\theta_0}^{(1)}$, we focus on $\gamma(\theta_0)$, which is defined as Eq.~(\ref{gamma_suppl}). For clarity, we write $\gamma(\theta_0)$ in the case of starting before an isentropic expansion (compression) stroke as $\gamma_{\text{exp}}$ ($\gamma_{\text{com}}$). Here, we consider $\gamma_{\text{exp}}$ as an example. Figure \ref{fig:C} shows the covariance matrix $C_{ij} \equiv C(t_i,\, t_j)$, which appears in the integrand of the expression of $\gamma(\theta_0)$ [Eq.~(\ref{gamma_suppl})]. For $\gamma_{\text{exp}}$, the elements of $C_{ij}$ enclosed by the dashed lines contribute, so that $\gamma_{\text{exp}}$ can be written as
\begin{align}
	\gamma_{\text{exp}} & = \frac{(C_{31}+C_{42}) + 2 C_{32} }{\text{Var}\big[W_{\text{exp}}^{(1)} \big] }  \nonumber\\
	& = a\frac{\phi_r^2+ (1-2 e_c^{-1}) }{ (1-2e_c) \phi_r^2 + 1} .
\end{align}
Similarly, $\gamma_{\text{com}}$ can be written as
\begin{equation}
	\gamma_{\text{com}} = a\frac{(1-2 e_h^{-1}) \phi_r^2 + 1}{\phi_r^2 + (1-2e_h)} .
\end{equation}
Thus the uncertainty $\Delta^{\infty}$ for infinite cycles is given by
\begin{equation}
	\Delta^{\infty}  = \Delta_{ \text{exp} }^{(1)} \frac{1+\gamma_{\text{exp}} }{1-a} = 2\frac{(1-2e_c+e_ce_h)\phi_r^2 + (1-2e_h+e_ce_h)}{(1-e_ce_h)(\phi_r-1)^2} .\label{eq:Dinfty_suppl}
\end{equation}
Therefore, from Eqs.~(\ref{D1_exp_suppl}), (\ref{D1_com_suppl}), and (\ref{eq:Dinfty_suppl}), the uncertainties $\Delta_{ \text{exp} }^{(1)} $, $\Delta_{ \text{com} }^{(1)}$, and $\Delta^{\infty}$ depend on the three parameters: $e_h$, $e_c$, and $\phi_r$. In addition, for $n$ cycles, we obtain
\begin{align}
	\Delta^{(n)}_{\text{exp}} & =  \bigg( 1 - \frac{s_n}{n} \bigg) \Delta^{\infty} + \frac{s_n}{n} \Delta^{(1)}_{\text{exp}} \notag \\
	& = 2 \left( 1 - \frac{1-(e_c e_h)^n}{n(1-e_c e_h)} \right) \frac{(1-2e_c+e_ce_h)\phi_r^2 + (1-2e_h+e_ce_h)}{(1-e_ce_h)(\phi_r-1)^2} + 2  \frac{1-(e_c e_h)^n}{n(1-e_c e_h)} \frac{(1-2e_c)\, \phi_r^2 + 1}{(\phi_r-1)^2}
\end{align}
and 
\begin{equation}
	\Delta^{(n)}_{\text{com}} = 2 \left( 1 - \frac{1-(e_c e_h)^n}{n(1-e_c e_h)} \right) \frac{(1-2e_c+e_ce_h)\phi_r^2 + (1-2e_h+e_ce_h)}{(1-e_ce_h)(\phi_r-1)^2} + 2  \frac{1-(e_c e_h)^n}{n(1-e_c e_h)} \frac{\phi_r^2 + (1-2e_h)}{(\phi_r-1)^2}.
\end{equation}

\bigskip
\subsection{E: Constraint on $e_h$, $e_c$, and $\phi_r$: derivation of Eq.~(12)}

	We consider the protocol starting from the adiabatic compression with $\lambda (t) = \lambda_h$ for $0<t<\tau_h$, $\lambda (t) = \lambda_c$ for $\tau_h<t<\tau$, $T(t) = T_h$ for $0<t<\tau_h$, and $T(t) = T_c$ for $\tau_h<t<\tau$, where $\lambda (t) = \lambda (t+\tau)$ and $T(t) = T(t+\tau)$. From Eq.~\eqref{phi_suppl}, the correlation function $\phi_{22} \equiv \phi(\tau,\tau) = \phi(0,0)$ is given by
	\begin{align}
		\phi_{22} &= 2\mu \frac{\displaystyle \int_{0}^{\tau} e^{2f(s)}T(s)\, \mathrm{d}s}{e^{2f(\tau)}-1}  \notag \\
		&= \frac{ 1}{1-e_c e_h}\left[ \dfrac{T_c}{\lambda_c} (1-e_c) + \dfrac{T_h}{\lambda_h} (1-e_h) e_c \right] .
	\end{align}
	Here, we have used the following results: $\exp[2f(t)] = \exp(2\mu\lambda_h t)$ for $0<t<\tau_h$, $\exp[2f(t)] = e_h^{-1}\exp[2\mu\lambda_c (t-\tau_h)]$ for $\tau_h<t<\tau$, $T(0^+)=T(\tau_h^-)=T_h$, and $T(\tau_h^+)=T(\tau^-)=T_c$. 
	In the same way, the correlation function $\phi_{11} \equiv \phi(\tau_h,\tau_h)$ is given by
	\begin{align}
		\phi_{11} & =e^{-[2f(\tau_h)]}\, 2\mu \left[ \int_0^{\tau_h} e^{2f(s)}T(s)\, \mathrm{d}s + \frac{\displaystyle \int_0^{\tau}e^{2f(s)}T(s)\, \mathrm{d}s}{e^{2f(\tau)}-1} \right] \notag \\
		& = 2\mu \frac{\displaystyle \int_{\tau_h}^{\tau+\tau_h} e^{2f(s)-2f(\tau_h)}T(s)\, \mathrm{d}s}{e^{2f(\tau)}-1}  \notag \\
		&= \frac{ 1 }{1-e_c e_h} \left[ \dfrac{T_c}{\lambda_c} (1-e_c) e_h + \dfrac{T_h}{\lambda_h} (1-e_h) \right].
	\end{align}
	
	The ratio $\phi_r \equiv \phi_{11}/\phi_{22}$ reads
	\begin{equation}
		\phi_r = \frac{R (1-e_c) e_h + (1-e_h) }{R (1-e_c) + (1-e_h)e_c} ,
	\end{equation}
	which gives a constraint on $e_h$, $e_c$, and $\phi_r$ for the Otto engine: 
	\begin{equation}
		\frac{(1-e_h)(1-\phi_r e_c)}{(1-e_c)(\phi_r- e_h)} = R \label{R_suppl}.
	\end{equation}

\bigskip

\subsection{F: The role of $e_h$, $e_c$, and $\phi_r$ on the power}

In the main paper, we have discussed about the correlation-enhanced stability in the Brownian Otto engine, and we have found that the uncertainty of work depends only on the three parameters: $e_h$, $e_c$, and $\phi_r$. Here, we show that the mean value of power for given $e_h$, $e_c$, and $\phi_r$ can reach arbitrary value for $0<e_c<1$ and $0<e_h<1$. In addition, we give a detailed calculation of the power plotted in Figs.~4(b)--4(e).

Power of the Brownian Otto engine is given by \cite{eta2}
\begin{equation}
	P = \frac{ \left(1-{\lambda_c}/{\lambda_h} \right) \left[ \left({\lambda_c}/{\lambda_h}\right) T_h-T_c \right]}{\left( {\lambda_c}/{\lambda_h} \right) \left(\tau_h+\tau_c \right)}\frac{\sinh \left(\mu \lambda_c \tau_c \right)\sinh \left(\mu \lambda_h \tau_h\right)}{\sinh \left(\mu \lambda_c \tau_c + \mu \lambda_h \tau_h \right)}, \label{P_Otto}
\end{equation}
which depends on six parameters: $T_h$, $T_c$, $\lambda_h$, $\lambda_c$, $\tau_h$, and $\tau_c$. At any point in Fig.~2(a), only three parameters are given: $e_h=\exp(-2\mu\lambda_h\tau_h)$, $e_c=\exp(-2\mu\lambda_c\tau_c)$, and $\phi_r$. Here, $\phi_r$ can be substituted by $R=T_c\lambda_h/(T_h\lambda_c)$ from Eq.~$(12)$. Therefore, we can still choose arbitrary positive values for the remaining three of the six parameters. For example, when $\lambda_h\tau_h$, $\lambda_c\tau_c$, and $R$ are given, we can still choose arbitrary positive values for $T_h$, $T_c$, and $\lambda_h$ with $T_c/T_h < R$. Here, the constraint on $T_c/T_h$ is from the fact that $\eta = 1-\lambda_c/\lambda_h = 1-(T_c/T_h)/R >0$ for the Otto engine. In this case, we can rewrite Eq.~\eqref{P_Otto} as follows:
\begin{equation}
	P = -\mu T_h \lambda_h \frac{ R-{T_c}/{T_h} }{ \ln e_h + (R{T_h}/{T_c}) \ln e_c} \frac{(1-R)}{R} \frac{(1-e_c)(1-e_h)}{1-e_c e_h} . \label{P_Otto2}
\end{equation}
Since there is no bound on $T_h$ or $\lambda_h$ in our discussion, power can take any positive value. Figures~4(b)--4(e) show the power as a function of $e_c$ and $e_h$ for given values of $T_h$, $T_c$, $\lambda_h$, and $\phi_r$ obtained from Eq.~\eqref{P_Otto2} and \eqref{R_suppl}.

\subsection{G: Correlation-enhanced stability for the Carnot cycle}	

We discuss the effect of intercycle correlation in the Carnot cycle with the model by Schmiedl and Seifert \cite{Schmiedl}. We calculate the uncertainty of work output at maximum power for the Carnot cycle, and show that the negative intercycle correlations can be obtained in some regions of the parameter space for the Carnot cycle as well. 

\begin{figure}[h]
	\includegraphics[width=0.5 \columnwidth]{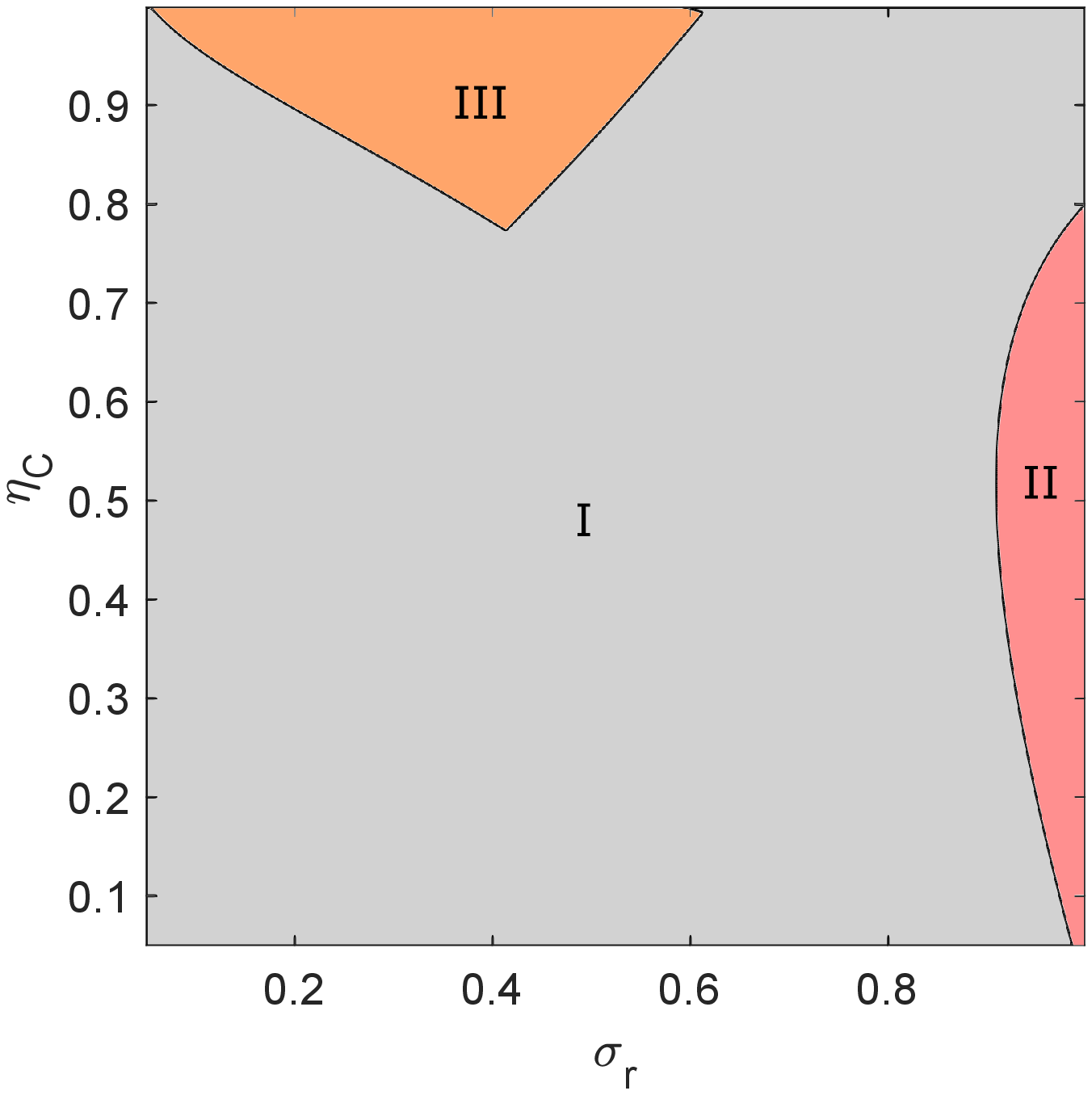}
	\centering
	\caption{Mapping out the regions of the correlation-enhanced stability for the Carnot cycle with the model by Schmiedl and Seifert \cite{Schmiedl}. $\Delta^{\infty}$ is compared to $\Delta_{\theta_0}^{(1)}$ with the starting point $t_0 = 0^+$, $\tau_h^-$, $\tau_h^+$, or $\tau^-$	corresponding to the node in the Carnot cycle. In regions \uppercase\expandafter{\romannumeral2} and \uppercase\expandafter{\romannumeral3}, $\Delta^{\infty}$ is smaller than $\Delta_{\theta_0}^{(1)}$ for any of the four choices of $t_0$. 
	}
	\label{fig:Schmiedl}
\end{figure}

In this model, the working substance is an overdamped Brownian particle trapped in a harmonic oscillator potential. According to Ref.~\cite{Schmiedl}, the cycle consists of two adiabatic jumps and two isothermal strokes.
The durations of the adiabatic strokes are negligible and those of the hot and cold isothermal strokes are $\tau_h$ and $\tau_c$, respectively. During the hot (cold) isothermal stroke, $0<t<\tau_h$ ($\tau_h<t<\tau=\tau_h+\tau_c$), the water temperature $T_h$ ($T_c$) is constant. The heat engine protocol $\lambda(t)$ is optimized to yield the maximum work output with given boundary values of the variance of the particle position $x$: $ \sigma_{\text{min}} \equiv \sigma(0) = \sigma(\tau)$ and $\sigma_{\text{max}} = \sigma(\tau_h)$, where $\sigma \equiv \langle x^2 \rangle - \langle x \rangle^2$. 
Since the power is nonzero at each moment, $\dot{W} \neq 0$, the correlation of work for the Carnot cycle is more complicated than that for the Otto cycle considered in the main paper. From our numerical result, we find that the uncertainty of work at maximum power depends only on the ratio $\sigma_r \equiv \sigma_{\text{min}}/\sigma_{\text{max}}$ and $\eta_C \equiv 1 - T_c/T_h$. Figure~\ref{fig:Schmiedl} shows that there are some regions (regions \uppercase\expandafter{\romannumeral2} and \uppercase\expandafter{\romannumeral3}) in the parameter space where we have the correlation-enhanced stability.

\end{document}